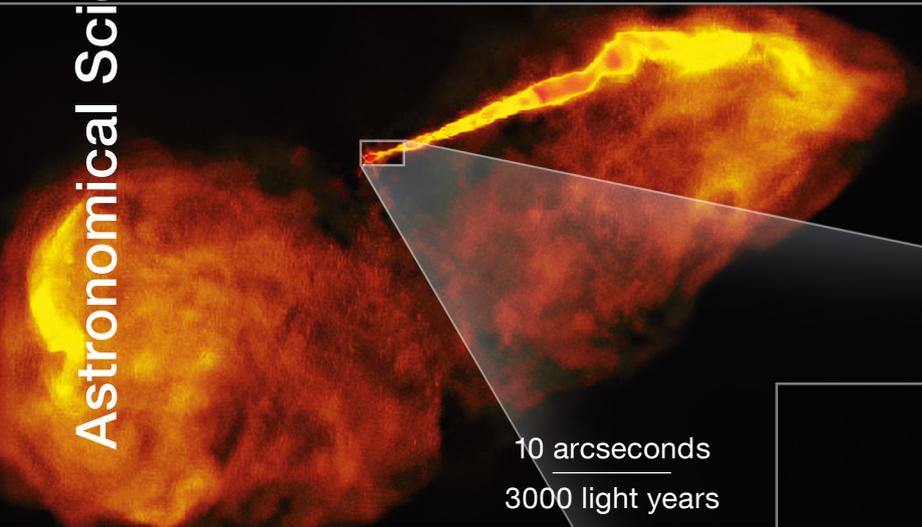

VLA – 1.5 GHz

10 arcseconds
3000 light years

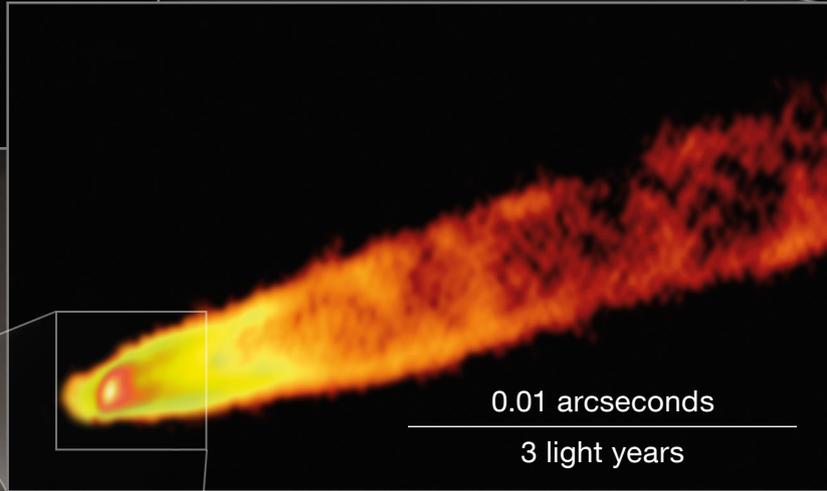

VLBA – 43 GHz

0.01 arcseconds
3 light years

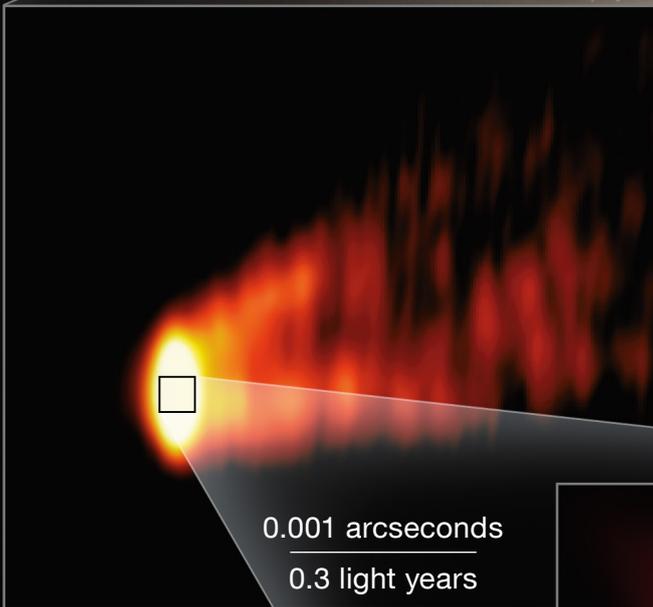

GMVA – 86 GHz

0.001 arcseconds
0.3 light years

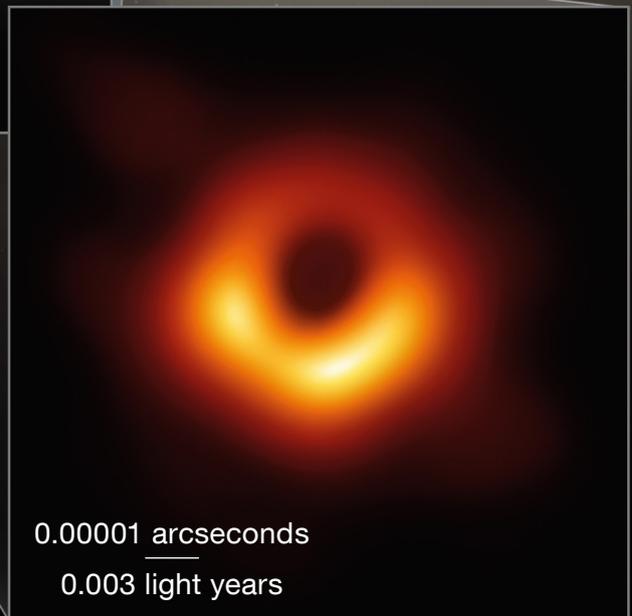

EHT – 230 GHz

0.00001 arcseconds
0.003 light years

The centre of the giant elliptical galaxy M87 seen at spatial resolution scales spanning six orders of magnitude. The detailed structure of the relativistic jet is revealed by observations at different radio wavelengths using several interferometric facilities, zooming into the supermassive black hole imaged by the EHT collaboration.

EHT Collaboration/M. Kornmesser/ESO

# First M87 Event Horizon Telescope Results and the Role of ALMA

Ciriaco Goddi[1,2]
Geoff Crew[3]
Violette Impellizzeri[4]
Iván Martí-Vidal[5,6]
Lynn D. Matthews[3]
Hugo Messias[4]
Helge Rottmann[7]
Walter Alef[7]
Lindy Blackburn[8]
Thomas Bronzwaer[1]
Chi-Kwan Chan[9]
Jordy Davelaar[1]
Roger Deane[10]
Jason Dexter[11]
Shep Doeleman[8]
Heino Falcke[1]
Vincent L. Fish[3]
Raquel Fraga-Encinas[1]
Christian M. Fromm[12]
Ruben Herrero-Illana[18]
Sara Issaoun[1]
David James[8]
Michael Janssen[1]
Michael Kramer[7]
Thomas P. Krichbaum[7]
Mariafelicia De Laurentis[19,20]
Elisabetta Liuzzo[21]
Yosuke Mizuno[12]
Monika Moscibrodzka[1]
Iniyan Natarajan[10]
Oliver Porth[14]
Luciano Rezzolla[12]
Kazi Rygl[21]
Freek Roelofs[1]
Eduardo Ros[7]
Alan L. Roy[7]
Lijing Shao[17,7]
Huib Jan van Langevelde[13,2]
Ilse van Bemmel[13]
Remo Tilanus[1,2]
Pablo Torne[15,7]
Maciek Wielgus[8]
Ziri Younsi[16,12]
J. Anton Zensus[7]
on behalf of the Event Horizon Telescope collaboration

[1] Department of Astrophysics, Institute for Mathematics, Astrophysics and Particle Physics (IMAPP), Radboud University, Nijmegen, the Netherlands
[2] Leiden Observatory—Allegro, Leiden University, Leiden, the Netherlands
[3] Massachusetts Institute of Technology Haystack Observatory, Westford, USA
[4] Joint ALMA Observatory, Vitacura, Santiago de Chile, Chile
[5] Onsala Space Observatory, Chalmers University of Technology, Sweden
[6] Department of Astronomy and Astrophysics/Astronomical Observatory, University of Valencia, Spain
[7] Max-Planck-Institut für Radioastronomie (MPIfR), Bonn, Germany
[8] Center for Astrophysics | Harvard & Smithsonian, Cambridge, USA
[9] Steward Observatory and Department of Astronomy, University of Arizona Tucson, USA
[10] Centre for Radio Astronomy Techniques and Technologies, Department of Physics and Electronics, Rhodes University, Grahamstown, South Africa
[11] Max-Planck-Institut für Extraterrestrische Physik, Garching, Germany
[12] Institut für Theoretische Physik, Goethe Universität, Frankfurt am Main, Germany
[13] Joint Institute for VLBI ERIC (JIVE), Dwingeloo, the Netherlands
[14] Anton Pannekoek Institute for Astronomy, University of Amsterdam, the Netherlands
[15] Instituto de Radioastronomía Milimétrica, IRAM, Granada, Spain
[16] Mullard Space Science Laboratory, University College London, Dorking, UK
[17] Kavli Institute for Astronomy and Astrophysics, Peking University, Beijing, China
[18] ESO
[19] Dipartimento di Fisica "E. Pancini," Universitá di Napoli "Federico II", Naples, Italy
[20] INFN Sez. di Napoli, Compl. Univ. di Monte S. Angelo, Naples, Italy
[21] INAF–Istituto di Radioastronomia, Bologna, Italy

In April 2019, the Event Horizon Telescope (EHT) collaboration revealed the first image of the candidate supermassive black hole (SMBH) at the centre of the giant elliptical galaxy Messier 87 (M87). This event-horizon-scale image shows a ring of glowing plasma with a dark patch at the centre, which is interpreted as the shadow of the black hole. This breakthrough result, which represents a powerful confirmation of Einstein's theory of gravity, or general relativity, was made possible by assembling a global network of radio telescopes operating at millimetre wavelengths that for the first time included the Atacama Large Millimeter/submillimeter Array (ALMA). The addition of ALMA as an anchor station has enabled a giant leap forward by increasing the sensitivity limits of the EHT by an order of magnitude, effectively turning it into an imaging array. The published image demonstrates that it is now possible to directly study the event horizon shadows of SMBHs via electromagnetic radiation, thereby transforming this elusive frontier from a mathematical concept into an astrophysical reality. The expansion of the array over the next few years will include new stations on different continents — and eventually satellites in space. This will provide progressively sharper and higher-fidelity images of SMBH candidates, and potentially even movies of the hot plasma orbiting around SMBHs. These improvements will shed light on the processes of black hole accretion and jet formation on event-horizon scales, thereby enabling more precise tests of general relativity in the truly strong field regime.

## Supermassive black holes and their shadows: a fundamental prediction of general relativity

Black holes are perhaps the most fundamental and striking prediction of Einstein's General Theory of Relativity (GR), and are at the heart of fundamental questions attempting to unify GR and quantum mechanics. Despite their importance, they remain one of the least tested concepts in GR. Since the 1970s, astronomers have been accumulating indirect evidence for the existence of black holes by studying the effects of their gravitational interaction with their surrounding environment. The first such evidence came from the prototypical high-mass X-ray binary Cygnus X-1, where a star orbits an unseen compact object of ~ 15 solar masses, apparently feeding on material from its stellar companion at only 0.2 au. More evidence has come from studies of the Galactic Centre, where ~ 30 stars have been tracked in tight, fast orbits (up to 10 000 km s$^{-1}$) around a radio point source named Sagittarius A* or Sgr A* (Gillessen et al., 2009), practically ruling out all mechanisms responsible for their motions, except for a black hole with a mass of about four million solar masses.

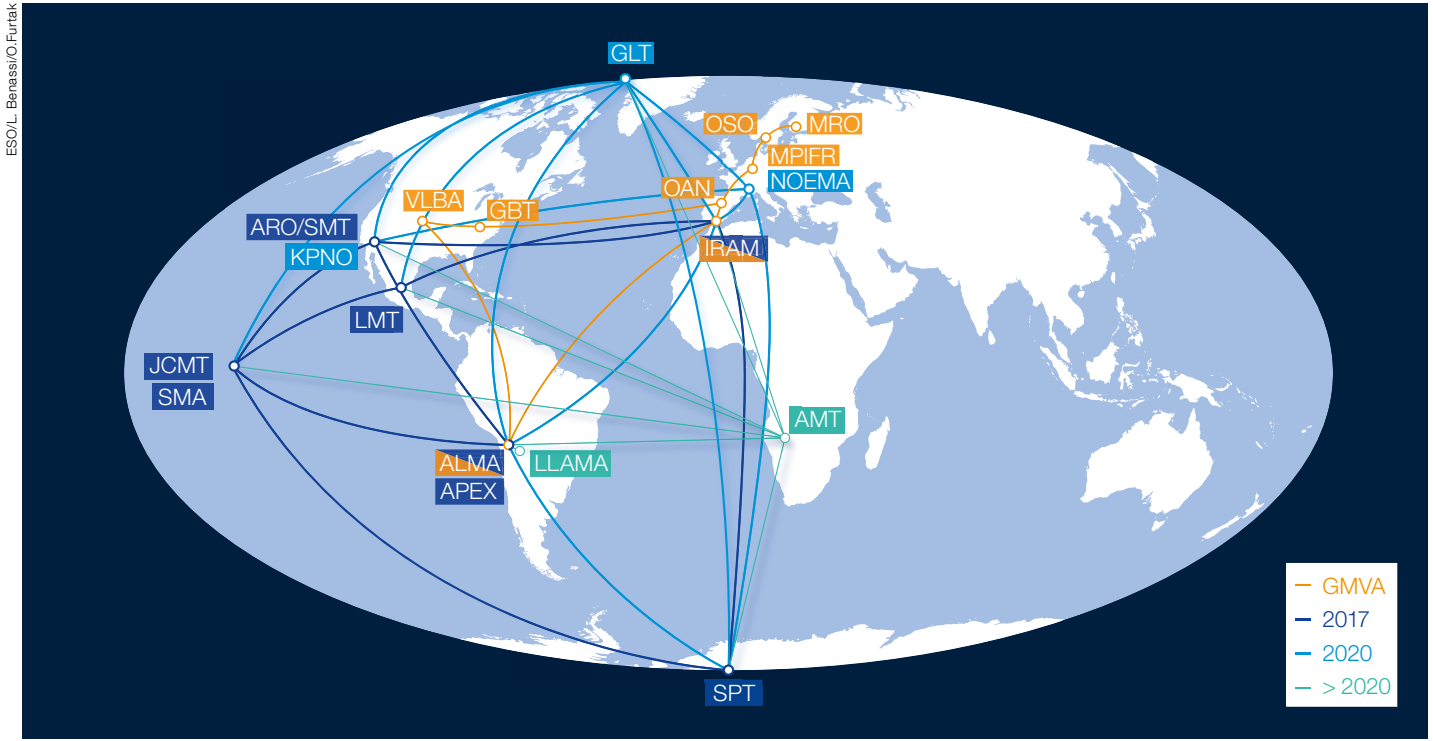

Figure 1. Locations of the participating telescopes of the Event Horizon Telescope (EHT; shown in blue) and the Global mm-VLBI Array (GMVA; shown in yellow) during the 2017 global VLBI campaign. Additional telescopes that will observe in 2020 are shown in light blue; the GLT also joined in the campaign conducted in 2018. Planned telescopes that may join the EHT in the future are shown in green.

Perhaps the most compelling evidence came in 2015, with the detection by the advanced Laser Interferometer Gravitational-Wave Observatory (LIGO) of gravitational waves: ripples in space-time produced by the merger of two stellar-mass black holes (Abbot et al., 2016). Despite this breakthrough discovery, there was until very recently no direct evidence for the existence of an event horizon, the defining feature of a black hole and a one-way causal boundary in spacetime from which nothing (including photons) can escape. On 10 April 2019, the EHT provided the very first resolved images of a black hole, demonstrating that they are now observable astrophysical objects and opening a new and previously near-unimaginable window onto black hole studies.

In order to conduct tests of GR using astrophysical black holes, it is crucial to observationally resolve the gravitational sphere of influence of the black hole, down to scales comparable to its event horizon. For a non-rotating black hole, the radius of the event horizon is equal to its Schwarzschild radius:
$R_{Sch} = 2\, GM_{BH}/c^2 = 2\, r_g$,
where $r_g$ is the gravitational radius, $M_{BH}$ is the black hole mass, G is the gravitational constant, and c is the speed of light. The angular size, subtended by a non-rotating BH with diameter $2\, R_{Sch}$ is:
$\theta_{Sch} = 2\, R_{Sch}/D \approx 40\, (M_{BH}/10^6\, M_\odot)(kpc/D)$
in microarcseconds (µas), where the black-hole mass is expressed in units of one million solar masses and the black hole's distance (D) is in kiloparsecs. For stellar-mass black holes (with masses of a few to tens of solar masses), $\theta_{Sch}$ lies well below the resolving power of any current telescope. SMBHs, which are thought to reside at the centre of most galaxies, are millions to billions of times the mass of the Sun, but as they are located at much greater distances, their apparent angular sizes are also generally too small to be resolved using conventional observing techniques. Fortunately, there are two notable exceptions: Sgr A* and the nucleus of M87.

Sgr A* and the nucleus of M87: the "largest" black hole shadows in our sky

Sgr A*, at the centre of our own Galaxy, hosts the closest and best constrained candidate SMBH in the Universe. With a mass of 4.15 million solar masses and at a distance of 26 400 light years or 8.1 kpc (Gravity collaboration et al., 2019), this SMBH is a factor of a million times larger than any stellar mass black hole in the Galaxy and at least a thousand times closer than any other SMBH in other galaxies. The second-best candidate is found in the nucleus of the giant elliptical galaxy M87, the largest and most massive galaxy within the local supercluster of galaxies in the constellation of Virgo. Located 55 million light years from the Earth (or 16.8 Mpc), it hosts a black hole of 6.5 billion solar masses. Therefore, even though M87 is ~ 2000 times as distant, it is ~ 1500 times as massive as Sgr A*, yielding a (slightly) smaller but comparable angular size of the black hole shadow on the sky. Owing to the combination of their masses and proximity, both Sgr A* and the nucleus of M87 subtend the largest angular size on the sky among all known SMBHs ($\theta_{Sch} \approx 20$ and 15 µas, respectively). This makes Sgr A* and M87 the two most suitable sources for studying the accretion process and jet formation in SMBHs, even enabling tests of GR at horizon-scale resolution.

The "shadow" of a black hole

Although by definition black holes cannot be seen, we can detect light which passes very close to the event horizon before escaping, allowing us to see what is around the black hole.

So what would a black hole actually look like if we could observe it? David Hilbert began calculating the bending of light around a Schwarzschild (non-rotating) black hole in 1917. Bardeen (1973) subsequently calculated the geometrical properties of a rotating black hole's silhouette against a bright background (an orbiting star). Although the likelihood of a black hole passing in front of a star is very small, black holes never appear "naked" in astrophysical environments since their extreme gravitational fields will pull and compress matter from their surroundings, eventually forming a disc of luminous plasma. Luminet (1979) performed simulations of a black hole surrounded by a geometrically thin, optically thick, accretion disc. Falcke, Melia and Agol (2000) demonstrated that an accreting black hole embedded in a plasma that is optically thin at millimetre wavelengths (like the plasma expected to surround Sgr A*) would produce a bright ring of emission with a dim "shadow" cast by the black hole event horizon in its interior. They suggested that such a shadow might be detectable towards the Galactic Centre using the technique of very long baseline interferometry (VLBI) at millimetre wavelengths[a].

The shadow and ring are caused by a combination of light bending and photon capture at the event horizon. The size scale of the emission ring is set by the photon capture radius $R_c$. For a non-rotating Schwarzschild black hole, $R_c = \sqrt{27}\, r_g \sim 2.5\, R_{Sch}$. The factor of ~ 2.5 comes from gravitational lensing, which increases the radius of the photon ring with respect to the Schwarzschild radius as seen from the observer, resulting in an angular diameter on the sky of ~ 50 and ~ 40 µas (as viewed from the Earth) for Sgr A* and M87, respectively. Although very small, this angular size can now be resolved by the VLBI technique at millimetre wavelengths using the EHT.

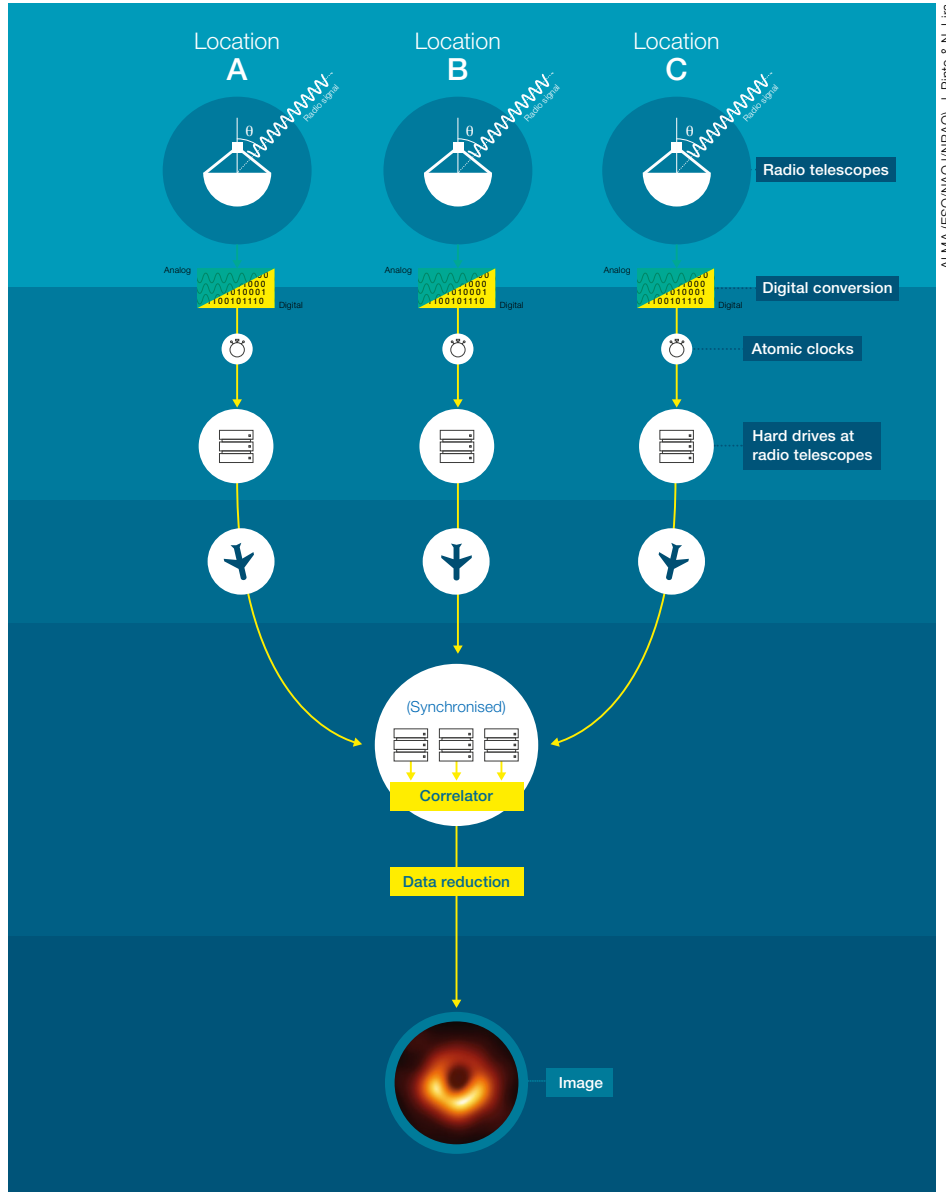

Imaging black holes with the Event Horizon Telescope (EHT)

The VLBI technique at millimetre wavelengths

For VLBI to work, a network of radio telescopes spread across different continents

Figure 2. A schematic diagram of the VLBI technique. Radio wave signals collected by individual antennas are converted from analogue to digital and recorded onto hard disks together with the time-stamps provided by extremely precise atomic clocks at each station. In the 2017 campaign, a total of about 4000 TB of recorded data was obtained. The data were shipped from each station to a central location where a supercomputer (the correlator) combined the signals between all pairs of antennas (synchronised using the timing information at each station). The output of the correlator is hundreds of gigabytes, which is further reduced during data calibration down to tens to hundreds of megabytes. The end product of the VLBI data processing is an astronomical image.

(see, for example, Figure 1) must observe the same source at exactly the same time and in the same frequency band. Individual antennas record their signals (plus time stamps from very precise atomic clocks) onto computer hard disks which are then shipped to a central location, where a supercomputer (called a correlator) combines (cross-correlates) the signals between all pairs of antennas, synchronising them using the recorded timing information from each station. Figure 2 is a diagram illustrating the data acquisition and processing path in a VLBI experiment.

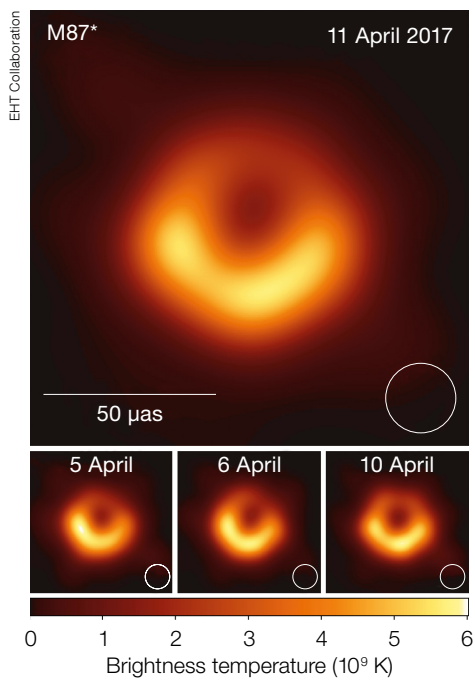

Figure 3. Image of the supermassive black hole M87* as obtained with the EHT (on four different days) in April 2017. Top panel: EHT image of M87* from observations on 11 April as a representative example of the images collected during the 2017 campaign. The angular resolution of the observation (20 μas) is shown in the lower right corner. The image is shown in units of brightness temperature. North is up and east is to the left. Bottom panels: similar images taken on different days showing the stability of the ring structure across the observing week.

The achievable angular resolution for an interferometer is given by $\theta \sim \lambda/B$ (in radians), where $\lambda$ is the observed wavelength and $B$ is the maximum distance between the telescopes (or baseline). Hence, higher frequencies (shorter wavelengths) and longer baselines provide the highest resolving power. At 1.3 mm (corresponding to a radio frequency of 230 GHz), Earth-diameter VLBI baselines achieve an angular resolution as fine as 20 μas, which is sufficient to resolve the shadow of both Sgr A* and M87. Therefore, the VLBI technique effectively mimics a virtual telescope with the size of the Earth.

While VLBI is well-established at centimetre wavelengths, its extension to wavelengths as short as 1.3 mm only began in the 1990s (for example, Padin et al., 1990; Krichbaum et al., 1997; Doeleman & Krichbaum, 1999). Challenges at shorter wavelengths include the reduced aperture efficiency and small diameter of radiotelescopes, increased noise in radio receiver electronics, higher atmospheric opacity, and above all, stronger distortion effects on the wavefronts from water vapour in the troposphere, which limits the phase coherence to only a few seconds. By 2003, several active galactic nuclei (AGN) had been detected on intercontinental baselines between Pico Veleta (Spain) and the Heinrich Hertz Telescope (HHT – Arizona, USA) at wavelengths of both 1 and 2 mm (Krichbaum et al., 2004; Doeleman et al., 2005).

Formation of the EHT project

Early pathfinder experiments (Krichbaum et al., 1998) detected Sgr A* with the baseline between Pico Veleta and the Plateau de Bure Interferometer, but the resolution was insufficient to probe horizon scales. In the mid-2000s, a focused effort to boost sensitivity through increased bandwidth led to the development of fully digital VLBI backends, with the goal of intercontinental 1.3-mm VLBI of Sgr A*. These systems were deployed at sites in Arizona, California, and Hawai'i, and event-horizon-scale structures were detected in both Sgr A* and M87 (Doeleman et al., 2008, 2012). These precursory scientific results motivated a strategy aimed at building a global 1.3-mm VLBI array capable of imaging the shadows of the SMBHs in both Sgr A* and M87, and spurred on the formation of the EHT project, which was proposed during the US 2010 Decadal Survey (Doeleman et al., 2009a)[b].

Since then, the EHT collaboration (or EHTC)[1] has grown to include over 250 members representing ~ 60 institutes, operating in over 20 countries/regions. The key elements of the roadmap towards an imaging array were the addition of new sites to better sample the Fourier plane, and the improvement in sensitivity needed to detect weak signals on short timescales (Doeleman et al., 2010). Two technological developments were crucial for the latter: (1) improvement in the observing bandwidth by increasing recording rates — over the last 10 years, EHT data rates increased from 4 to 64 Gb per second (Gbps); and (2) development of phased-array systems to combine the collecting area of existing connected-element (sub)millimetre interferometers, which has led to the inclusion in the EHT of ALMA and the Submillimeter Array (SMA) and the future incorporation of the NOrthern Extended Millimeter Array (NOEMA).

Phasing ALMA: turning ALMA into a giant single-dish VLBI station

ALMA is the most sensitive (sub)millimetre-wavelength telescope ever built. It consists of 54 12-metre and 12 7-metre antennas located on the Chajnantor plateau in the Atacama desert in Chile, the highest, driest (accessible) desert on the Earth, and it ordinarily operates as a connected-element interferometer. Although the implementation of a VLBI mode was not part of the baseline project, the desirability of phasing the entire array for VLBI had been recognised (Wright et al., 2001; Shaver, 2003) and some of the architecture needed to sum signals from all ALMA antennas was built into the ALMA correlator (Escoffier et al., 2007).

Motivated by the prospect of using ALMA for horizon-scale observations of supermassive black holes (Doeleman et al., 2009a, 2010), the case for phasing ALMA was renewed. An international team, led by Doeleman at MIT Haystack Observatory, proposed an ALMA Phasing Project (APP) to the US National Science Foundation. The APP was accepted by the ALMA Board in 2011 and was completed in 2018 thanks to an international effort with contributors from the USA, Europe, and Asia.

The heart of the APP is a beamformer system, which electronically combines the collecting area of ALMA by aligning the signals from individual ALMA antennas in phase to form a coherent sum signal. Currently, up to 43 12-metre antennas are used for the phased array, but the number may be smaller depending on the array configuration and the weather conditions. This effectively turns ALMA into a giant virtual single dish (hereinafter called "phased ALMA"), and is equivalent to adding a ~ 70-metre dish to existing mm-VLBI arrays. In order to phase ALMA successfully and incorporate it as a VLBI station, the APP had to add several new hardware and software components. These included an optical fibre link system to transport the phased-sum signal from the ALMA Array Operations Site (at an altitude of 5100 m) to the ALMA Operations Support Facility (at 2900 m), where a set of Mark 6 VLBI recorders was installed. ALMA's original rubidium clock was replaced with a more precise hydrogen maser (required to properly tune and synchronise the signals in the VLBI array). Numerous software enhancements were also required, including the implementation of an ALMA VLBI Observing Mode (VOM) and a phase solver to adjust the phase of each of the ALMA antennas during observations to allow coherent summation of their signals. The initial scope of the APP effort and implementation can be found in Doeleman et al. (2010); the final working implementation is described in Matthews et al. (2018) and Goddi et al. (2019).

A broad science case for the use of the ALMA Phasing System was assembled by the international community in white papers[c] (Fish et al., 2013; Tilanus et al., 2014). Starting in 2016, VLBI as an observing mode was made available to the astronomy community through the normal ALMA proposal system, with an expected maximum time allocation of ~ 5% of the total ALMA observing time.

Figure 4. The supermassive black hole at the centre of M87. Left: The black hole feeds on a swirling disc of glowing plasma (shown in red), driving a powerful relativistic jet across several thousands of light years (shown in grey; simulation by Davelaar et al. 2018). Bottom right: Approaching the black hole, gravity is so strong that light is severely bent, creating a bright (almost circular) ring. The north–south asymmetry in the emission ring is produced by relativistic beaming and Doppler boosting (matter in the bottom part of the image is moving toward the observer) and is mediated by the black hole spin (which is pointing away from Earth and rotating clockwise). Gravitational lensing magnifies the apparent size of the black hole's event horizon into a larger dark shadow. The emission between the photon ring and the event horizon is due to emitting plasma either in the accretion flow and/or at the footprint of the jet (this emission is generally too dim to be detected by the EHT; see Younsi et al., in preparation for detail). Top right: While the EHT can zoom in very close to the event horizon, down to scales of only 0.01 light years (or 3.7 light days), i.e., a region comparable to the size of our Solar System, the relativistic jet (extended across several thousand light years) can be probed using ALMA intra-baselines, recorded during the EHT observations (greyscale image; Goddi et al. 2019 and in prep.).

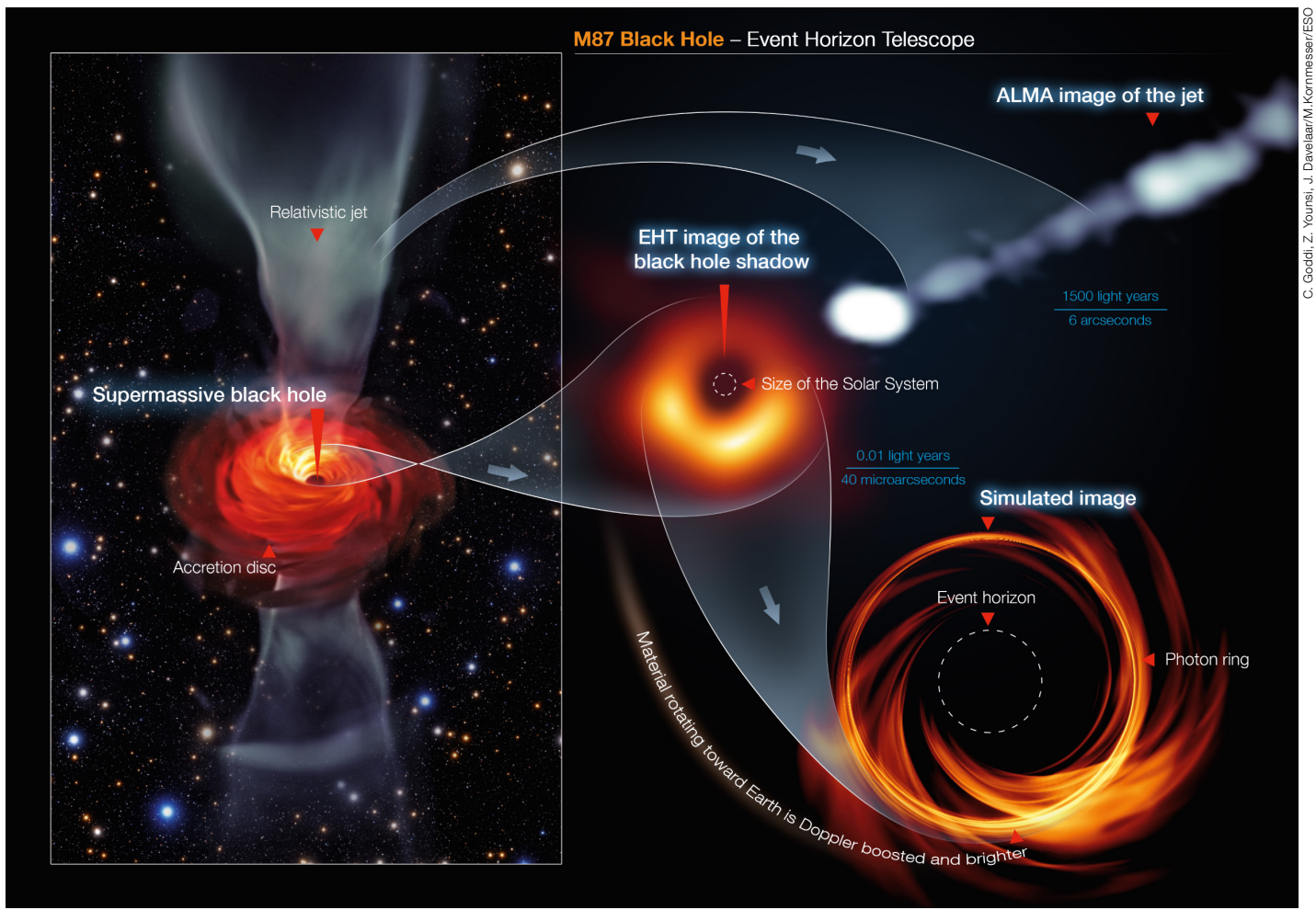

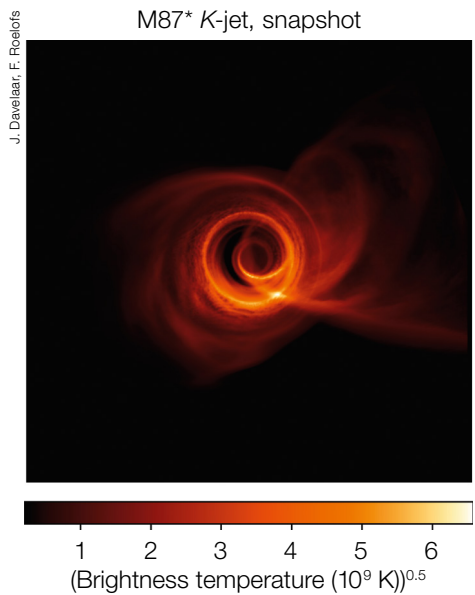

Figure 5. Black hole model for M87* used for the image reconstructions shown in Figure 6. This specific model has a relatively bright jet footprint appearing in front of the photon ring and a more extended jet emission extending towards west. Note the bright knot to the south-west at the point where the jet sheath crosses the photon ring in projection (see Davelaar et al., 2019).

## Impact of ALMA in the EHT array

Owing to the combination of a large effective aperture, its central location in the VLBI array, excellent typical atmospheric conditions and ultra-low noise receivers, the addition of ALMA drastically changed the overall capabilities of the global EHT array, boosting the achievable signal-to-noise ratio (SNR) of VLBI baselines by more than an order of magnitude with respect to the first horizon-scale detections (Doeleman at al., 2008, 2012). More specifically, the inclusion of ALMA in the EHT provides three key advantages: a boost in sensitivity; improved baseline coverage; and an accurate measurement of the absolute flux-density scale and polarisation fractions of EHT targets (using standard ALMA interferometric data). We expand on these characteristics below.

I. The median thermal noise of non-ALMA baselines is 7 mJy, and 0.7 mJy in ALMA baselines; for M87 this corresponds to SNR > 10 on non-ALMA baselines and > 100 on ALMA baselines. Therefore, the addition of ALMA into the EHT array greatly facilitates detections, especially for weak signals (for example, long baseline length or bad weather). Using ALMA as a highly sensitive reference station has enabled critical corrections for ionospheric and tropospheric distortions at the other EHT sites (see EHTC et al., 2019c for details).

II. ALMA has a central location in the EHT array (see Figure 1), and is therefore essential for the baseline coverage and image fidelity. Even though ALMA and the Atacama Pathfinder Experiment (APEX) are extremely close geographically, ALMA's superior sensitivity allows the EHT to detect signals with the required 10-second integration times between all baselines, i.e., to find VLBI fringe solutions, which has a dramatic impact on the imaging capability of the EHT (see Figure 6).

III. VLBI observations with ALMA also provide connected-element interferometric data, which are archived, as with any standard ALMA project, and are available to the user in the ALMA archive (after the appropriate proprietary period). As outlined in Goddi et al. (2019), the calibration of such interferometric data allows one to determine the absolute amplitude calibration of the co-located sites ALMA–APEX in physical flux-density (i.e, Jy). This, in turn, allows us to bootstrap source fluxes and calibrate longer baselines across the entire array (i.e., network calibration). In addition, since VLBI observations are always performed in full-polarisation mode (in order to supply input to the polarisation conversion process at the VLBI correlators — see Martí-Vidal et al., 2016; Goddi et al., 2019), the ALMA full-polarisation interferometric datasets can be used to derive mm-wavelength emission and polarisation properties of each target observed by the EHT on arcsecond scales (Goddi et al., in preparation).

## The first global VLBI campaigns with ALMA

Phased ALMA joined the EHT array for the first time in April 2017, performing VLBI observations with an array capable of imaging its key science targets. The global array included eight telescopes in six different geographical sites: the South Pole Telescope (SPT), the Arizona Radio Observatory's Submillimeter Telescope (SMT), the Large Millimeter Telescope Alfonso Serrano (LMT) in Mexico, the IRAM 30-metre telescope in Spain, the SMA and the James Clerk Maxwell Telescope (JCMT) in Hawai'i, and APEX and ALMA in Chile. These telescopes provided baseline lengths up to 10 700 km towards M87, resulting in an array with a resolution of ~ 20 μas (details are provided in EHTC et al., 2019b).

Besides the EHT, which operates at a wavelength of 1.3 mm (i.e., a frequency of 230 GHz, ALMA Band 6), complementary VLBI observations with ALMA were also conducted at 3.5 mm or 86 GHz (ALMA Band 3) in concert with the Global mm-VLBI Array (GMVA)[3], which consists of up to 18 telescopes located in Europe, North America, and Asia. Figure 1 displays the geographical locations of all the participating telescopes in the EHT and the GMVA in 2017 (plus additional telescopes that joined after or plan to join in the near future).

The EHT 2017 science observing campaign was scheduled for April when Sgr A* and M87 are night-time sources and tropospheric conditions tend to be the best, averaged over all sites in the array. At that time, ALMA was in a more compact configuration as required for phased array operations. About 40 EHT astronomers travelled to four continents and drove to mountaintops to carry out the observations, which included six science targets: the primary EHT targets, Sgr A* and M87, and the AGN targets, 3C 279, OJ 287, Centaurus A, and NGC 1052[d]. Weather is a crucial factor for VLBI observations at mm wavelengths, which is why the EHT uses flexible observing schedules with windows that are about twice as long as the approved number of observing nights. An array-wide go/no-go decision was made a few hours before the start of each observing night, based on the weather conditions and predictions at each site, as well as technical readiness at each of the participating telescopes. Observations were triggered over a 10-night window 5–14 April 2017 — on 5, 6, 7, 10, and 11 April. During the whole campaign, the weather was good to excellent at most stations. In addition to favourable weather conditions, the VLBI-specific technical setup and operations at all sites were successful, which resulted in fringe detections across the entire array.

Figure 6. Image reconstructions from synthetic data generated using the M87* model displayed in Figure 5 as input. Each panel shows a reconstruction from simulated observations with a different array: 2017 EHT array (left column); planned 2020 array (middle column); and a future EHT array including AMT in Namibia, and LLAMA in Argentina (right column). A comparison between the top row (with ALMA) and the middle row (without ALMA) highlights the importance of ALMA, demonstrating that the shadow feature can only be recovered if ALMA is part of the array. A comparison between the middle row and the bottom row (where APEX is also excluded) clearly demonstrates that, even with a single-dish telescope in the same geographical location of ALMA, it is not possible to recover an image with sufficiently high fidelity to discern the shadow. Although with future arrays the quality of the reconstructed image will improve and it will be possible to recover new features (for example, the jet), the addition of new stations cannot fully compensate for the loss of ALMA (middle and right columns).

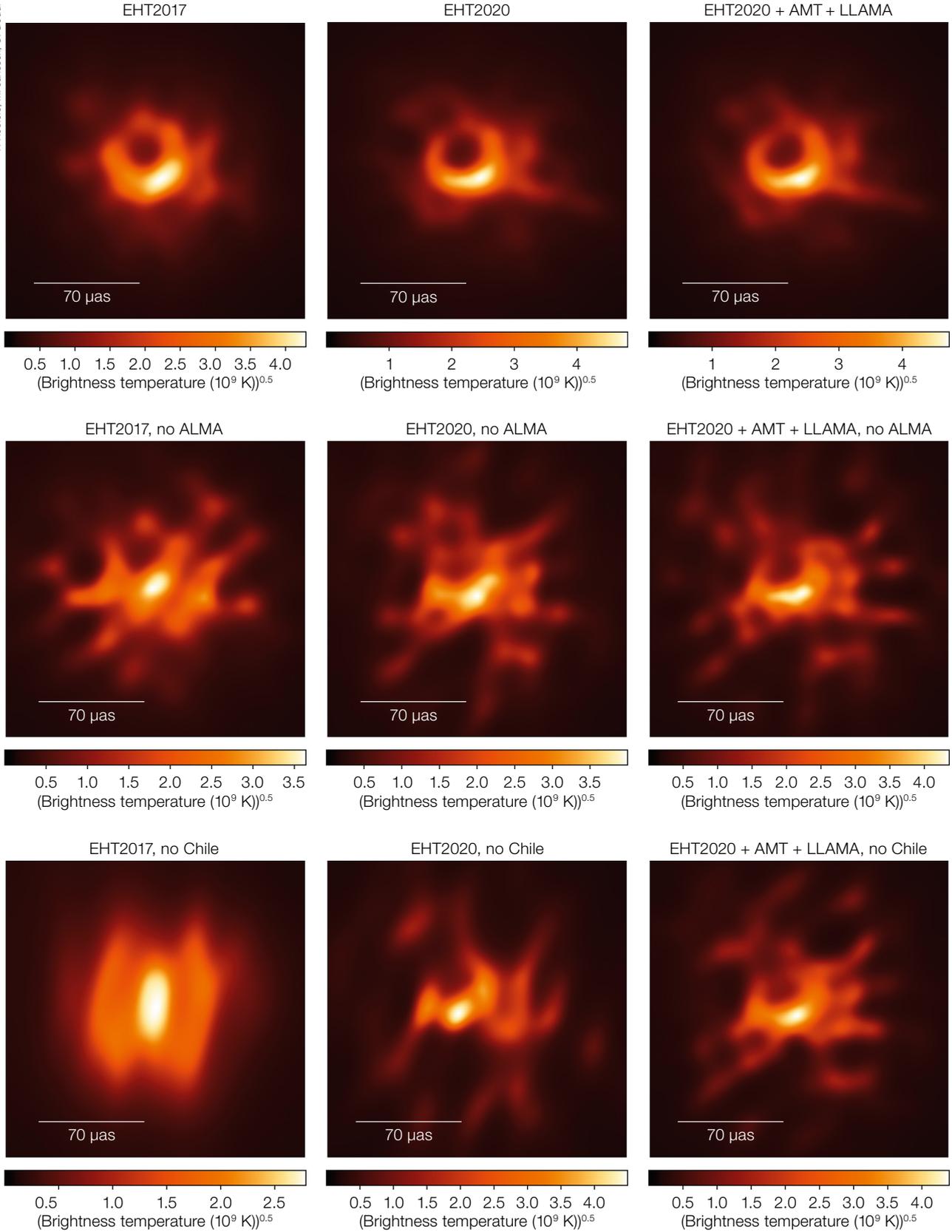

By the end of the campaign, 96 disk modules were used, each containing eight helium-filled hard disks (of either six or eight terabyte [TB] capacity), corresponding to more than 5 petabytes (PB) of (removable) storage; about 4 PB of data were eventually recorded in total. Given this huge volume of data, the observing campaign VLBI data could not be transferred over the internet, but were shipped from each remote station to the two EHT correlator centres for processing (the shipping took at least several days, and many months in the case of the South Pole telescope).

Path to the image

The EHT data were correlated at the MIT Haystack Observatory in Westford, USA and at the MPIfR in Bonn, Germany. Three independent data calibration pipelines and three imaging pipelines, each using a different software package and associated methodology, were used for the data processing in order to produce images independently. This approach encourages the data analysis and science teams to minimise their biases in terms of both methodology and human decision making; further details of this are provided in EHTC et al. (2019c,d). Multiple independent analyses were also performed in order to verify the results (EHTC et al., 2019e,f). After about two years of dedicated work by many dozens of EHT scientists in multiple working groups (from instrument through data processing to theory), the collaboration was finally ready to communicate our breakthrough to the world.

The breakthrough: first image of a black hole shadow

The first image of a black hole was published by the EHT Collaboration on 10 April 2019 in a series of six scientific publications (EHTC et al., 2019a,b,c,d,e,f). The announcement and the historic image were released worldwide in six simultaneous press conferences — in Washington, Brussels, Santiago, Shanghai, Taipei and Tokyo — with additional satellite events in Rome, Madrid, Munich, Leiden and Nijmegen, amongst others. The core of M87, as imaged with the EHT, was renamed M87*, in line with the name of the black hole candidate at the centre of the Galaxy, Sgr A*.

The most striking feature of the image (displayed in Figure 3) is a bright circular ring with an asymmetric brightness distribution and a dark region at its centre, which identifies the black hole shadow. The ring reveals the curvature of spacetime due to the extreme gravitational field around a SMBH, which bends light around it, creating an almost circular shadow at its centre. In fact, GR predicts the shadow to be circular to within a few percent, whereas alternative theories of gravity predict distorted, non-circular shapes (Younsi et al., 2016; also see Figure 7 of Goddi et al., 2017).

The ring has a measured diameter of 42 ± 3 µas and the central brightness depression has a contrast ratio > 10:1. The measured angular size, assuming a distance of 16.8 Mpc (EHTC et al., 2019f), implies a black hole mass of $M = (6.5 ± 0.7) × 10^9 M_⊙$, or 6.5 billion times the mass of the Sun (consistent with one earlier mass measurement — Gebhardt et al., 2011). To convert the measured diameter of the ring into the mass of the black hole, the radiating plasma around the black hole was modelled with general relativistic calculations spanning a wide range of possible accretion states (see next subsection). By tracing the peak of the emission in the ring we can determine the shape of the image, which is close to circular with an axial ratio 4:3 (corresponding to a 10% deviation from circularity). The emission in the ring is asymmetric and is brighter in the south, which can be explained as relativistic beaming of plasma rotating (close to the speed of light) in the clockwise direction around the black hole as seen by the observer (i.e., the bottom part of the emission ring is Doppler-boosted towards the Earth). Based on our modelling and information on the inclination angle of the relativistic jet (observed on larger scales), we derive the sense of rotation of the black hole to be in the clockwise direction, i.e., the spin axis of the black hole points away from us.

A number of elements reinforce the robustness of the result. The data analysis used four independent data sets taken on four different days (spanning a one-week observing window) in two separate frequency bands (centered at 227 and 229 GHz). The top part of Figure 3 shows an image of M87* on 11 April, while the bottom panels show similar images from three different days. The diameter and width of the ring remain stable and the image features are broadly consistent across all four observing days, except the position angle of the bright part in the asymmetric azimuthal profile, which varies in the range 150–200 degrees measured from north towards the east between the first two days and the last two days.

Overall, the size, circularity, asymmetry, and brightness contrast of the observed image are consistent with the shadow of a "Kerr" black hole as predicted by GR and provide the strongest evidence to date of the existence of SMBHs in the nuclei of external galaxies.

Modelling and physical interpretation of the black hole image

The appearance of M87* has been modelled using 3D general-relativistic magnetohydrodynamic (GRMHD) simulations, which provide the physical conditions of the plasma and magnetic field surrounding the black hole. GR ray-tracing radiative-transfer (GRRT) codes then take this GRMHD simulation data as input and calculate the black hole's appearance from the emitted radiation field. Approximately 60 000 simulated images were produced in the process (see EHTC et al., 2019e). Figure 4 showcases the main components of the M87 SMBH and their characteristic scales by comparing observed and simulated images (for one specific set of models). In particular, in the simulation to the left (which combines emission at wavelengths of 7, 3, and 1 mm; see Davelaar et al., 2018 for details), one can see that the SMBH is embedded in an accretion flow and powers a bipolar relativistic jet.

Zooming in closer to the centre (the simulation to the right in Figure 4 shows emission at 1 mm; Younsi et al., in preparation), hot magnetised plasma orbiting and accreting onto the black hole creates the familiar emission ring structure around the event horizon. As stated earlier, the size of the ring is set by the photon capture radius: photons approaching the black hole with an impact parameter $b < R_c$ are captured and disappear into the black hole; photons with $b > R_c$ escape to infinity; photons with $b = R_c$ are captured on an unstable circular orbit and produce the so-called lensed photon ring.

While the EHT can resolve down to scales of 0.01 light years (or 3.7 light-days), i.e., a region comparable to the size of our Solar System, the jet extends to much larger scales across several thousand light years and can be probed using shorter baselines and/or lower frequencies. In Figure 4 (top right), we show an image of the M87 jet, which extends across about 20 arcseconds, corresponding to 5000 light years, obtained at 1.3 mm using ALMA interferometric data (with maximum baseline lengths of only a few hundred metres) acquired simultaneously with the EHT observations (Goddi et al., 2019 and in preparation). The science section image (p. 24) showcases a montage of images of the M87 relativistic jet observed at several radio wavelengths with multiple interferometers at progressively higher angular resolution overlaid on the HST optical image: the VLA at 21 cm, VLBA at 7 mm, GMVA at 3 mm, and EHT at 1.3 mm.

Implications of the black hole shadow on tests of GR and complementarity with LIGO

Simulated images can be used to test basic properties of black holes as predicted in GR (for example, Psaltis et al., 2015), or in alternative theories of gravity (Younsi et al., 2016; Mizuno et al., 2018). They can also be used to test alternatives to black holes (Olivares et al., 2019). We estimate a deviation from circularity of < 10 %, so we can set an initial limit on relative deviations from GR. Although it is difficult to rule out alternatives to black holes in GR — a shadow can be produced by any compact object with unstable circular photon orbits (Mizuno et al., 2018) — we can readily exclude exotic alternatives to black holes, such as naked singularities or wormholes, which predict much smaller shadows than we have measured, whereas others like boson stars and gravastars need to be analysed with more care (Olivares et al., 2019); also see EHTC et al. (2019e) for further details.

It is worth pointing out that the EHT result provides a new way to study black hole spacetimes and is complementary to the detection experiments of gravitational waves from merging stellar-mass black holes with LIGO/Virgo (Abbott et al., 2016). There are at least two main complementary aspects between gravitational-wave and electromagnetic observations of black holes:
1. Since EHT targets SMBHs and LIGO mainly targets stellar-mass black holes, combining measurements from both methods we can test whether one of the most fundamental properties of black holes in GR, that their size scales linearly with mass, actually holds over eight orders of magnitude.
2. Gravitational wave experiments cannot rely on the possibility of multiple and repeated measurements of the same source, whereas the EHT can be used to measure the shadow shape of M87 with ever increasing precision, leading to progressively better constraints on black hole parameters and their spacetime.

Importance of ALMA in imaging M87

The most straightforward way to visualise how the loss of specific stations changes the baseline coverage and sensitivity, thereby affecting the resulting images, is by performing simulated observations. Instrument simulators were specifically built for the EHT to tie theoretical models to instrument measurements. In particular, they can generate realistic synthetic data, taking as input GRMHD model images, and performing synthetic observations using a specific EHT array and observing schedule (see EHTC et al., 2019d,e). As a demonstration, in Figure 5 we show a GRMHD simulation of the jet-launching region of M87 from Davelaar et al. (2019). This specific model has a relatively bright jet footprint appearing in front of the photon ring and a more extended jet emission extending towards west. We can then test how well the input image (the ground truth) can be recovered with the current EHT array and analyse the effect of adding new stations and/or excluding existing stations.

Figure 6 shows some examples of reconstructed images of M87, using the model shown in Figure 5 as the input model, produced with one specific EHT synthetic data generation pipeline (details are reported in Roelofs, Janssen and EHTC, submitted). For instance, using the EHT array and schedule that observed on 11 April 2017, the resulting simulated image (shown in the left panel, top row of Figure 6) is similar to the one actually observed by the EHT. The middle and bottom rows show simulated images without ALMA, which best showcase its importance by clearly demonstrating that the familiar ring structure cannot be reconstructed when ALMA is not part of the array. Although APEX shares the same geographical location as ALMA and therefore should provide similar baseline coverage, the quality of the reconstructed image is not sufficient to discern the ring when APEX is in the array and ALMA is excluded. These simulations clearly substantiate the need for ALMA's sensitivity, which allows for numerous and strong detections of weak signal[e].

By adding new stations, the quality of the reconstructed image improves and new features (for example, the jet) can be recovered (see middle and right columns in the top row of Figure 6), but these new stations cannot fully compensate for the loss of ALMA (see middle and right columns in the middle and bottom rows).

## Future directions

The first EHT image of M87 has provided very strong evidence for the existence of an event horizon and supports the notion of SMBHs being located at the centre of galaxies. SMBHs present a new tool to explore gravity at its most extreme limit and on a mass scale that was hitherto inaccessible. Ongoing analysis of existing data and future EHT observations will further help us understand the nature of black holes and will provide even more stringent tests of GR.

Future observations and detailed analysis of M87 data will explore the shape and time variability of the shadow more accurately. The EHT is in the process of studying the magnetised plasma around M87 in polarised light, which will allow us to investigate the mechanism by which black holes launch and power their relativistic jets.

As for Sgr A*, the mass-to-distance ratio is accurately measured from stellar orbits in the near-infrared (Gravity Collaboration et al., 2019), so measuring the shadow shape and diameter provides a null hypothesis test of GR (Psaltis et al., 2015). Since its mass is three orders of magnitude smaller than that of M87*, the dynamical timescales are minutes instead of days; therefore observing the shadow of Sgr A* will require accounting for this variability as well as the mitigation of scattering effects caused by the interstellar medium (Johnson, 2016). Time-dependent non-imaging analysis can potentially be used to track orbits of hot spots near the black hole (Broderick & Loeb, 2006; Doeleman et al., 2009b; Roelofs et al., 2017), as reported recently on the basis of interferometric observations in the near-infrared (Gravity Collaboration et al., 2018). Real-time movies may also become possible via interferometric dynamical imaging (Johnson et al., 2017). Time-domain studies and movies of black holes can then be used to study black hole accretion and to map the black hole spacetime, leading directly to measurements of black hole spin and tests of the "no hair" theorem (Broderick et al., 2014).

Although the focus of this article is the first EHT result from the 2017 campaign, it is noteworthy that enhancement of the EHT's capabilities over the coming years will bring more exciting scientific results. In 2018, the Greenland Telescope (GLT) joined the EHT (and GMVA) and VLBI observations were conducted as part of ALMA Cycle 5 (the analysis of these observations is still ongoing). In 2019, EHT observations were abandoned because of operational difficulties at a small number of key EHT sites. For 2020, observations are planned during ALMA Cycle 7 and will include new telescopes: Kitt Peak National Observatory (KPNO) in Arizona, and the NOEMA interferometer in France. These new stations will provide intermediate baselines ($\lesssim$ 1000 km) in Europe (NOEMA–IRAM-30m) and important short baselines in the USA (KPNO–SMT) ($\lesssim$ 100 km), thus further extending the baseline coverage for both M87 and Sgr A*. The possible future addition of the Africa Millimeter Telescope (AMT) in Namibia (Backes et al., 2016) and the Large Latin American Millimeter Array (LLAMA) in Argentina will add further baseline coverage, including the long baselines oriented east-west (AMT–ALMA) and the intermediate baseline LLAMA–ALMA (180 km). In particular, the addition of short/intermediate baselines of the order of a few hundred km, which are sensitive to extended emission from the jet on scales > 100 µas, may enable us to trace the jet down to the SMBH and directly image the jet launching. See the middle and right columns of Figure 6 to evaluate the impact of these new stations in recovering the jet structure; the location of the new stations is also displayed in Figure 1.

Higher-resolution images can be achieved by going to a shorter wavelength (0.87 mm or 345 GHz, i.e., ALMA Band 7). A future array that combines observations at both 1.3 and 0.87 mm will improve the imaging dynamic range, while multi-frequency VLBI would also open up spectral index and rotation measure studies.

In the more distant future, extending VLBI into space would provide the increased angular resolution necessary to image finer structures and dynamics near the black hole shadow (Fish et al., 2019; Palumbo et al., 2019; Roelofs et al., 2019). An order of magnitude increase in angular resolution would allow us to perform precision tests of GR and to measure parameters like black hole spin. While current terrestrial VLBI at 1.3 mm can resolve the black hole shadow only in Sgr A* and M87, adding satellites in space would significantly expand the range of sources that can be resolved on horizon scales. Combining ground-based VLBI at 0.87 mm with space-based VLBI at longer wavelengths would provide better matching beam sizes, which are important for spectral index and rotation measure studies. Studying nearby low-luminosity AGN could fill the gaps in black hole mass, accretion/jet power, and host galaxy type between Sgr A* and M87. Therefore all these developments will open up very exciting and new scientific possibilities in the coming decades.

Finally, if we were to discover a radio pulsar on a tight orbit (period < 1 year) around Sgr A*, this would allow us to measure the black hole properties (mass, distance and spin) more accurately than currently possible with orbiting stars targeted by the AO-assisted, two-object, multiple beam-combiner interferometric VLTI instrument, GRAVITY, leading to a clean test of the no-hair theorem (Psaltis, Wex & Kramer, 2016). The detection of a magnetar at a projected distance of 0.1 pc from Sgr A* (Eatough et al., 2013) suggests that finding a pulsar in a close orbit around Sgr A* should be possible, and the recent detection of the Vela pulsar with phased ALMA (Liu et al., 2019) opens up the possibility of pulsar searches with ALMA at high frequencies (where the effect of interstellar scattering is lower). The combination of the far-field measurements (100s–1000s $r_g$) based on pulsars and stars, with the near-field tests from imaging of black hole shadows (10s rg), has the power to reveal deviations from the Kerr metric and provide a fundamental test of GR (Goddi et al., 2017), potentially leading to a breakthrough in our understanding of physics in the strong gravity regime.

### Acknowledgements

The authors would like to acknowledge all the scientists, institutes, observatories and funding agencies who are part of and collectively support the EHT project. The APP was supported by a Major Research Instrumentation award from the National Science Foundation (NSF; award 1126433), an ALMA

North American Development Augmentation award, ALMA North America (NA) Cycle 3 and Cycle 4 Study awards, and an ALMA NA Cycle 5 Development award. The EHT project has been supported by multiple grants from many independent funding agencies, including the ERC Synergy Grant "BlackHoleCam: Imaging the Event Horizon of Black Holes" (Grant 610058) and several USA NSF grants (including AST-1310896, AST-1440254, and OISE-1743747). For the complete list of funding grants and acknowledgments please see EHT Collaboration et al. (2019a); they have not been reproduced here for reasons of space and readability. We gratefully acknowledge the support provided by the staff of the ALMA observatory.

This paper makes use of the following ALMA data: ADS/JAO.ALMA#2016.1.01154.V.

ALMA is a partnership of ESO (representing its member states), NSF (USA) and NINS (Japan), together with NRC (Canada), MOST and ASIAA (Taiwan), and KASI (Republic of Korea), in cooperation with the Republic of Chile. The Joint ALMA Observatory is operated by ESO, AUI/NRAO and NAOJ.

### Links

[1] The Event Horizon Telescope webpage: http://eventhorizontelescope.org/
[2] The Global mm-VLBI Array webpage: http://www3.mpifr-bonn.mpg.de/div/vlbi/globalmm/
[3] Astronet 2007: https://www.eso.org/public/archives/oldpdfs/Astronet_ScienceVision_lowres.pdf

### Notes

[a] We refer to Luminet (2019) for a comprehensive review of the history of early numerical simulations of black-hole imaging during the period 1972–2002 and Falcke (2017) for a review of past, current and future efforts to image black holes.
[b] The strategic importance of millimetre-VLBI for event-horizon-scale imaging of the Galactic Centre was also recognised in the European Science Vision for astronomy, Astronet 2007[3].
[c] In building the science case for phased ALMA, astronomers from all ALMA regions were asked for input at a variety of meetings in the USA and Europe, including an ESO workshop (see Falcke et al., 2012).
[d] In addition to being scientific targets, AGN observations are very important to facilitate the intensity and polarisation calibration of the entire VLBI array.
[e] APEX participation is nevertheless extremely important as it provides the ALMA–APEX short baseline (2.6 km), which allows the robust absolute calibration of visibility amplitudes (i.e., telescope sensitivities).

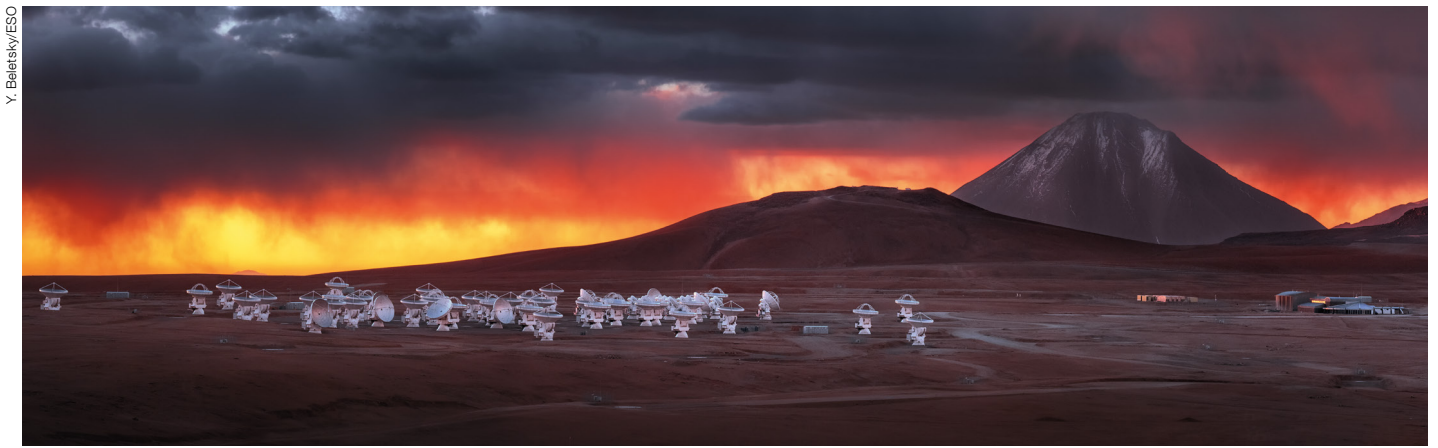

The groundbreaking ALMA array is composed of 66 giant antennas situated on the Chajnantor Plateau in the Chilean Andes.